\documentclass[final,5p,twocolumn]{elsarticle}

\usepackage{amssymb}
\usepackage{amsmath}

\usepackage[T1]{fontenc}
\usepackage[english]{babel}
\usepackage{subcaption}
\usepackage{adjustbox}
\usepackage{orcidlink}
\usepackage{flushend}

\newcommand{\csf}{$\gamma+e\rightarrow\gamma'+e'$}
\newcommand\BSTR{\rule[-1.2ex]{0pt}{0pt}} 

\journal{Physics Letters B}

\longmktitletrue
\begin{document}

\begin{frontmatter}

\title{Measurement of the Total Compton Scattering Cross Section between 6.5 and 11 GeV}
\affiliation[asu]{organization={Polytechnic Sciences and Mathematics, School of Applied Sciences and Arts, Arizona State University}, city={Tempe},  state={Arizona},  postcode={85287},  country={USA}}
\affiliation[athens]{organization={Department of Physics, National and Kapodistrian University of Athens}, city={Athens},  postcode={15771},  country={Greece}}
\affiliation[rub]{organization={Ruhr-Universit\"{a}t-Bochum, Institut f\"{u}r Experimentalphysik}, city={Bochum},  postcode={D-44801},  country={Germany}}
\affiliation[cmu]{organization={Department of Physics, Carnegie Mellon University}, city={Pittsburgh},  state={Pennsylvania},  postcode={15213},  country={USA}}
\affiliation[cua]{organization={Department of Physics, The Catholic University of America}, city={Washington},  state={D.C.},  postcode={20064},  country={USA}}
\affiliation[uconn]{organization={Department of Physics, University of Connecticut}, city={Storrs},  state={Connecticut},  postcode={6269},  country={USA}}
\affiliation[duke]{organization={Department of Physics, Duke University}, city={Durham},  state={North Carolina},  postcode={27708},  country={USA}}
\affiliation[fiu]{organization={Department of Physics, Florida International University}, city={Miami},  state={Florida},  postcode={33199},  country={USA}}
\affiliation[fsu]{organization={Department of Physics, Florida State University}, city={Tallahassee},  state={Florida},  postcode={32306},  country={USA}}
\affiliation[gwu]{organization={Department of Physics, The George Washington University}, city={Washington},  state={D.C.},  postcode={20052},  country={USA}}
\affiliation[giessen]{organization={Physikalisches Institut, Justus-Liebig-Universit\"{a}t Gie{\ss}en}, city={Gie{\ss}en},  postcode={D-35390},  country={Germany}}
\affiliation[glasgow]{organization={School of Physics and Astronomy, University of Glasgow}, city={Glasgow},  postcode={G12 8QQ},  country={United Kingdom}}
\affiliation[gsi]{organization={GSI Helmholtzzentrum f\"{u}r Schwerionenforschung GmbH}, city={Darmstadt},  postcode={D-64291},  country={Germany}}
\affiliation[ihep]{organization={Institute of High Energy Physics}, city={Beijing},  postcode={100049},  country={People’s Republic of China}}
\affiliation[iu]{organization={Department of Physics, Indiana University}, city={Bloomington},  state={Indiana},  postcode={47405},  country={USA}}
\affiliation[kurchatov]{organization={National Research Centre Kurchatov Institute}, city={Moscow},  postcode={123182},  country={Russia}}
\affiliation[lamar]{organization={Department of Physics, Lamar University}, city={Beaumont},  state={Texas},  postcode={77710},  country={USA}}
\affiliation[umass]{organization={Department of Physics, University of Massachusetts}, city={Amherst},  state={Massachusetts},  postcode={01003},  country={USA}}
\affiliation[mephi]{organization={National Research Nuclear University Moscow Engineering Physics Institute}, city={Moscow},  postcode={115409},  country={Russia}}
\affiliation[mau]{organization={Department of Physics, Mount Allison University}, city={Sackville},  state={New Brunswick},  postcode={E4L 1E6},  country={Canada}}
\affiliation[nsu]{organization={Department of Physics, Norfolk State University}, city={Norfolk},  state={Virginia},  postcode={23504},  country={USA}}
\affiliation[ncat]{organization={Department of Physics, North Carolina A\&T State University}, city={Greensboro},  state={North Carolina},  postcode={27411},  country={USA}}
\affiliation[uncw]{organization={Department of Physics and Physical Oceanography, University of North Carolina at Wilmington}, city={Wilmington},  state={North Carolina},  postcode={28403},  country={USA}}
\affiliation[odu]{organization={Department of Physics, Old Dominion University}, city={Norfolk},  state={Virginia},  postcode={23529},  country={USA}}
\affiliation[regina]{organization={Department of Physics, University of Regina}, city={Regina},  state={Saskatchewan},  postcode={S4S 0A2},  country={Canada}}
\affiliation[spring]{organization={Department of Mathematics, Physics, and Computer Science, Springfield College}, city={Springfield},  state={Massachusetts},  postcode={01109},  country={USA}}
\affiliation[jlab]{organization={Thomas Jefferson National Accelerator Facility}, city={Newport News},  state={Virginia},  postcode={23606},  country={USA}}
\affiliation[tomskpoly]{organization={Laboratory of Particle Physics, Tomsk Polytechnic University}, city={Tomsk},  postcode={634050},  country={Russia}}
\affiliation[tomsk]{organization={Department of Physics, Tomsk State University}, city={Tomsk},  postcode={634050},  country={Russia}}
\affiliation[union]{organization={Department of Physics and Astronomy, Union College}, city={Schenectady},  state={New York},  postcode={12308},  country={USA}}
\affiliation[vtech]{organization={Department of Physics, Virginia Tech}, city={Blacksburg},  state={Virginia},  postcode={24061},  country={USA}}
\affiliation[wjc]{organization={Department of Physics, Washington \& Jefferson College}, city={Washington},  state={Pennsylvania},  postcode={15301},  country={USA}}
\affiliation[wm]{organization={Department of Physics, William \& Mary}, city={Williamsburg},  state={Virginia},  postcode={23185},  country={USA}}
\affiliation[wuhan]{organization={School of Physics and Technology, Wuhan University}, city={Wuhan},  postcode={430072},  country={People’s Republic of China}}
\affiliation[yerevan]{organization={A. I. Alikhanyan National Science Laboratory (Yerevan Physics Institute)}, city={Yerevan},  postcode={0036},  country={Armenia}}
\author{F.~Afzal\orcidlink{0000-0001-8063-6719}\fnref{rub}}
\author{C.~S.~Akondi\orcidlink{0000-0001-6303-5217}\fnref{fsu}}
\author{M.~Albrecht\orcidlink{0000-0001-6180-4297}\fnref{jlab}}
\author{M.~Amaryan\orcidlink{0000-0002-5648-0256}\fnref{odu}}
\author{S.~Arrigo\fnref{wm}}
\author{V.~Arroyave\fnref{fiu}}
\author{A.~Asaturyan\orcidlink{0000-0002-8105-913X}\fnref{jlab}}
\author{A.~Austregesilo\orcidlink{0000-0002-9291-4429}\fnref{jlab}}
\author{Z.~Baldwin\orcidlink{0000-0002-8534-0922}\fnref{cmu}}
\author{F.~Barbosa\fnref{jlab}}
% ...

\begin{abstract}
The total cross section for Compton scattering off atomic electrons, \csf, was measured using photons with energies between 6.5 and 11.1~GeV incident on a $^9$Be target as part of the PrimEx-\textit{eta} experiment in Hall D at Jefferson Lab. This is the first measurement of this fundamental QED process within this energy range. The total uncertainties of the cross section, combining the statistical and systematic components in quadrature, averaged to 3.4\% across all energy bins. This not only demonstrates the capability of this experimental setup to perform precision cross-section measurements at forward angles but also allows us to compare with state-of-the-art QED calculations.
\end{abstract}


\end{frontmatter}

\section{Introduction}
\label{sec:introduction}

\begin{figure}[h]
    \centering
    \includegraphics[width=0.92\linewidth]{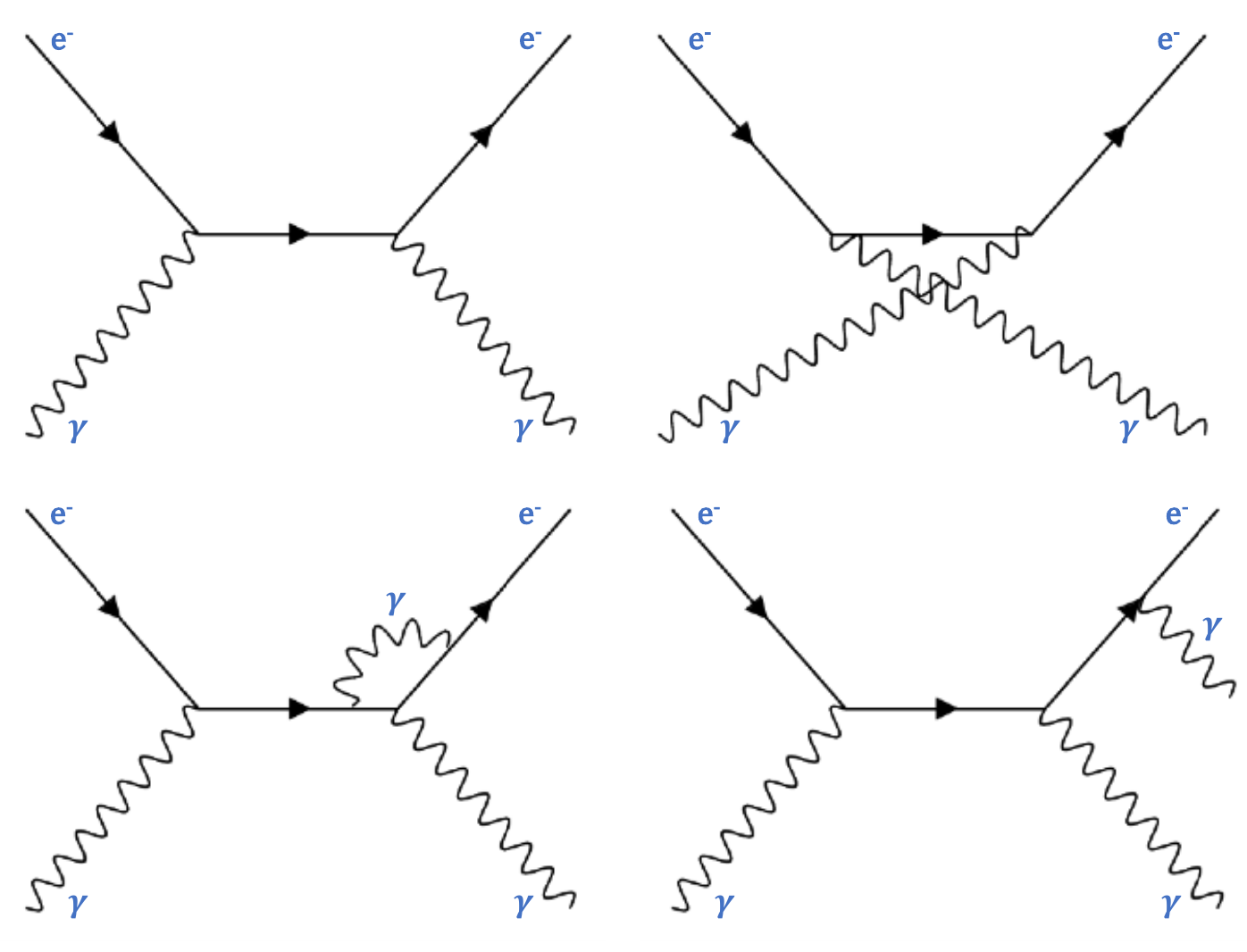}
    \caption{Top: The lowest-order Feynman diagrams for single Compton scattering. Bottom Left: Typical NLO radiative-correction diagram. Bottom Right: NLO double Compton scattering diagram.}
    \label{fig:feynman_diag}
\end{figure}

Compton scattering of photons off electrons $(\gamma + e \rightarrow \gamma' + e')$ is among the most fundamental processes within quantum electrodynamics (QED). 
The leading-order diagrams (top row of Fig.~\ref{fig:feynman_diag}) were first calculated by Klein and Nishina in 1929~\cite{Klein1929}, as well as Tamm in 1930~\cite{Tamm1930}. 
At Next-to-Leading Order (NLO), there are two classes of processes that introduce corrections to the leading-order Compton scattering cross section:
(i) the radiative process of emission and subsequent reabsorption of a virtual photon and (ii) the double Compton scattering process, both shown in the bottom row of Fig.~\ref{fig:feynman_diag}.
The first quantitative evaluations of these NLO corrections appeared in the 1950s~\cite{Brown1952,Mandl1952}, with progressively refined numerical techniques developed throughout the following two decades~\cite{Anders1966,Mork1971,Ram1971}.
More recently, in 2021, a closed-form analytical expression for the total cross section at NLO was obtained~\cite{Lee2021:PRL}.
These corrections are about 4\% of the total cross section at 10~GeV, with their relative impact increasing at higher photon energies.

Although QED has been established for nearly five decades, experimental results verifying the NLO corrections to the Compton scattering cross section for photon energies exceeding 1~GeV remain scarce.
The PrimEx-II experiment, completed in Hall B at Jefferson Lab, provided the first total cross-section measurement at the few-percent level precision~\citep{PrimExCompton}.
That result was obtained for only a narrow range of photon beam energies between 4.4 and 5.3~GeV, but it showed general agreement in magnitude and direction of the NLO corrections to the total cross section.
At higher energies, however, no results exist capable of testing the higher-order corrections.
In this paper, we present experimental results of the total Compton scattering cross section with photon energies between 6.5 and 11~GeV.
By extending to higher photon energies and covering a wider range, this measurement offers increased sensitivity to the NLO corrections and a more complete test of the theory.

In addition to bridging an important gap in experimental data, this result demonstrates for the first time the capability of this experimental setup to perform absolute cross section measurements with few-percent level precision. 
This is of particular importance for the PrimEx-\textit{eta} experiment, which plans to measure the $\eta$ meson radiative decay width via the Primakoff effect~\citep{PrimExEta}, and will use Compton scattering as a reference process to control common sources of systematic uncertainty and monitor experimental stability.

\section{Theoretical Background}
\label{sec:theory}

At Leading Order (LO), the differential cross section for Compton scattering is given by the Klein-Nishina formula~\cite{Klein1929}:
\begin{eqnarray}
 \frac{d\sigma_\text{LO}}{d\Omega} &=& \frac{r_{e}^{2}}{2}\left[1 + \gamma\left(1-\cos\theta_{\gamma}\right)\right]^{-2} \nonumber\\
 &\times& \left[1+\cos^{2}\theta_{\gamma} + \frac{\gamma^{2}\left(1-\cos\theta_{\gamma}\right)^{2}}{1 + \gamma\left(1-\cos\theta_{\gamma}\right)}\right],
 \label{eq:klein-nishina}
\end{eqnarray}
where $r_{e}$ is the classical electron radius, $\gamma = \frac{E_{\gamma}}{m_{e}c^{2}}$ is the Lorentz factor given by the ratio of the initial photon energy to the electron mass, and $\theta_{\gamma}$ is the scattering angle of the outgoing photon in the lab frame. 

At NLO, corrections are made to the cross section to account for loop contributions to $\gamma+e^{-}\rightarrow\gamma+e^{-}$ and the double Compton scattering effect: $\gamma+e^{-}\rightarrow\gamma+e^{-}+\gamma$. 
The contribution of the former to the differential cross section was first calculated by Brown and Feynman in 1951, and shown to contain an infrared divergence ~\cite{Brown1952}.
At the same time, they showed that the class of NLO corrections of double Compton scattering with the emission of a very soft secondary photon also contains an infrared divergence that exactly cancels the one coming from the loop diagrams.
This divergence was lifted by introducing a cutoff energy, $\omega_{2max} \ll m_{e}$, to define an upper limit of integration over the secondary (real) photon energies.
When combined with the loop diagrams of the same order, a single, physically meaningful correction can be determined for the total Compton scattering cross section.

A derivation of the differential cross section for double Compton scattering was first obtained by Mandl and Skyrme in 1952~\citep{Mandl1952}, and the numerical integration and evaluation of the total cross section was first carried out by Mork in 1971~\cite{Mork1971}. A more recent evaluation of the total cross section for double Compton scattering was calculated in Ref.~\cite{Lee2021:JHP}.
The full NLO corrections to the total Compton scattering cross section, combining the double Compton effect and the loop diagrams, were obtained in Ref.~\cite{Mork1971}, and the contribution from the emission of hard secondary photons (energy greater than $\omega_{2max}$) is given by Eq.~6.6 of that paper.
To summarize, the total cross section at NLO can be expressed as:
\begin{equation}
    \sigma_\text{NLO} = \sigma_\text{LO}\left(1 + \delta_\text{SV} + \delta_\text{DH}\right),
\end{equation}
where $\delta_\text{SV}$ is the contribution from the virtual photon loop diagrams together with soft double photon emission (SV: Soft + Virtual), and $\delta_\text{DH}$ is the contribution from emission of a hard secondary photon (DH: double, hard photon).
The contribution of $\delta_\text{SV}$ to the total cross section is negative, while the contribution of $\delta_\text{DH}$ is positive.

A program was adapted from the PrimEx-II experiment that used the BASES/SPRING Monte Carlo simulation package~\cite{Kawabata1995} to perform the numerical integration of the differential cross section at NLO based on the calculations of Ref.~\cite{Mork1971}.
The first step of this program (BASES) used a stratified sampling technique to perform the numerical integration of the LO cross section along with the Soft+Virtual correction over the photon scattering angle, and the numerical integration of the hard-photon double Compton process over four independent variables: the energy of each photon, and the opening and azimuthal angles between them.
Afterwards, the second part of the program (SPRING) generated Compton scattering events for simulation studies based on the probability information supplied by BASES.
The details of the numerical integration can be found in Ref.~\cite{PrimExCompton}.
Ultimately, the correction to the leading-order cross section by the NLO effects was found to be between $\sim$3.5\% for photons at 6.5~GeV and $\sim$4.0\% for photons at 11~GeV.

The total cross section calculation was validated by comparing it to a recently-obtained analytical expression for the NLO cross section given in Ref~\cite{Lee2021:PRL}. 
For all photon-energy bins used in this experiment, the cross section calculated by the event generator was in agreement with the analytical result to within 0.1\%.

\section{Experimental Details}

The Compton scattering process was measured using the GlueX spectrometer in Hall D at Jefferson Lab~\cite{GlueX2021}. 
An unpolarized bremsstrahlung photon beam was produced via the electrons from the Continuous Electron Beam Accelerator Facility (CEBAF) passing through a 10$^{-4}$ radiation length (RL) aluminum foil. 
The electrons were supplied in bunches, at a frequency of 249.5~MHz and an average beam current of 200~nA.
Downstream of the radiator, the recoiling electrons were deflected by a 6~m long dipole magnet and detected by an array of scintillating tagging counters to provide accurate energy and timing information of the bremsstrahlung photons.

A pair spectrometer (PS)~\cite{Barbosa2015} was used to determine the photon flux by detecting $e^+e^-$ pairs produced via photon conversions inside a 750~$\mu$m-thick beryllium foil positioned in the photon beam.
The resulting leptons were deflected by a dipole magnet and detected by scintillating detectors situated symmetrically with respect to the beamline, forming the electron and positron arms of the PS.
The detection acceptance of the PS for $e^+e^-$ pairs was measured from dedicated calibration data, in which a lead-tungstate (PbWO$_4$) total absorption counter was inserted into the photon beam.
These calibration data enabled a simultaneous measurement of the number of reconstructed electromagnetic pairs and the number of incident photons, thereby allowing for a precise determination of the PS acceptance. 
To facilitate the detection of individual beam photons, the calibration data were collected using a low-intensity photon beam, with a flux approximately 500 times lower than that used during nominal production runs.
A similar technique was employed by the PrimEx-II collaboration in Hall B at Jefferson Lab for their photon flux measurement~\cite{PrimExCompton}.
Roughly 10~m downstream of the PS, the bremsstrahlung photon beam impinged on a 5\% RL $^{9}$Be target, yielding an integrated luminosity of approximately 0.7~pb$^{-1}$ for tagged photons between 6.5 and 11~GeV.

\begin{figure}
    \centering
    \includegraphics[width=\linewidth]{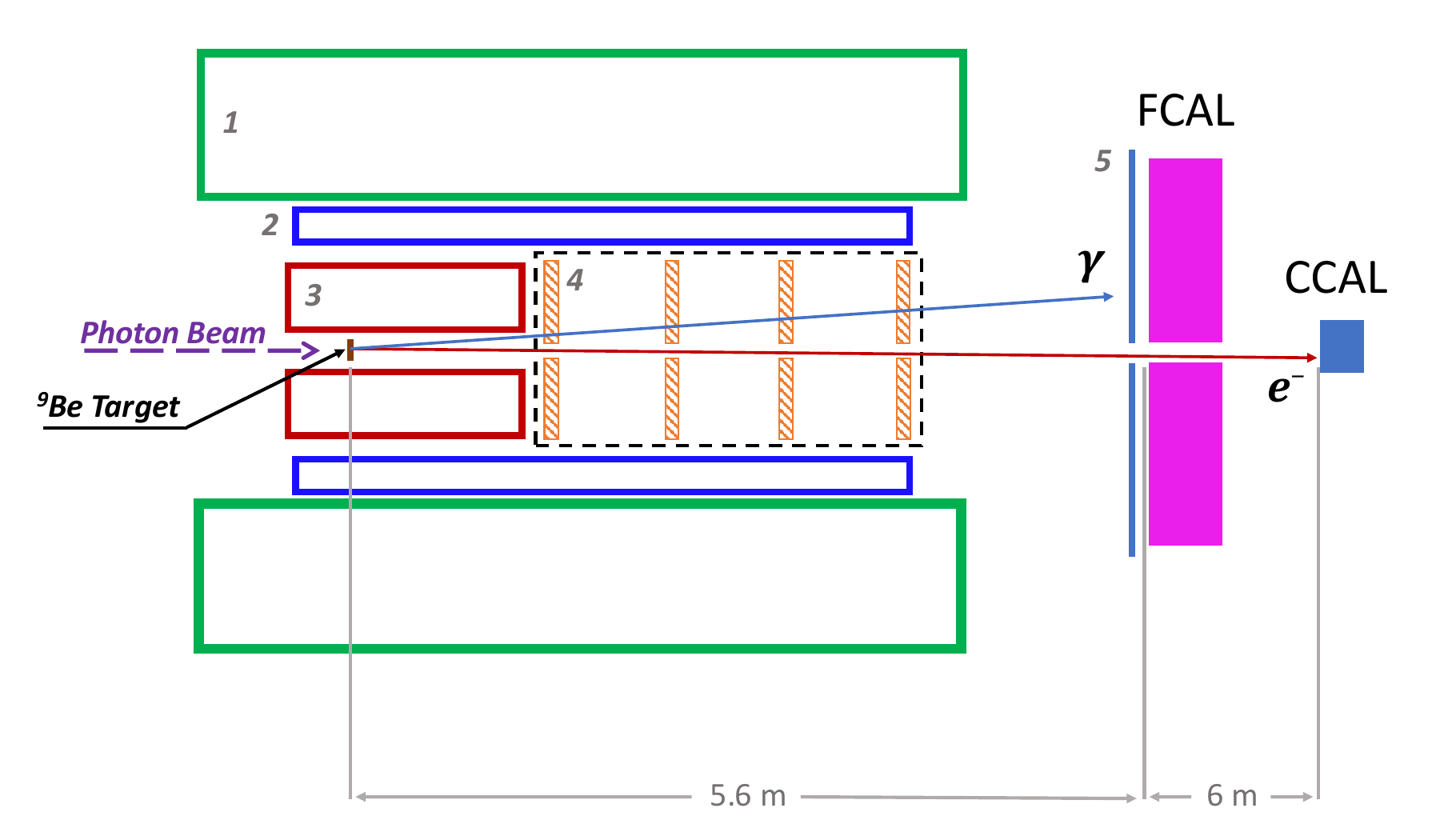}
    \caption{Detector layout for the PrimEx-\textit{eta} experiment. Numbered components include (1) solenoid magnet, (2) barrel electromagnetic calorimeter, (3) central drift chamber, (4) forward drift chambers, and (5) time-of-flight wall. The scattered photons and electrons from Compton scattering events were detected by the Forward Calorimeter (FCAL) and the Compton Calorimeter (CCAL). Not drawn to scale.}
    \label{fig:detector_layout}
\end{figure}

A schematic of the detector setup for the PrimEx-\textit{eta} experiment is shown in Fig.~\ref{fig:detector_layout}. 
Details on the detector components can be found in Ref.~\cite{GlueX2021}.
For the Compton scattering measurement reported in this article, the solenoid magnetic field was turned off to avoid perturbing the straight track of the scattered electron.
Without the magnetic field, the forward drift chambers also had to be turned off due to the high rates of electromagnetic background in the forward direction.
Consequently, no particle identification was performed in this analysis to distinguish the electron and photon on an event-by-event basis.

The Forward Calorimeter (FCAL) is an array of lead-glass modules located approximately 5.6~m downstream of the target center, and measures particles with (laboratory) scattering angles between 1$^{\circ}$ and 11$^{\circ}$.
A $12 \times 12$~cm$^{2}$ region in the central part of the FCAL is removed to allow the photon beam to pass through.
However, due to the kinematics of Compton scattering at $\sim$10~GeV photon energy, at least one of the final-state particles will always be emitted at an angle below 1$^{\circ}$ and pass through this beam hole.
In order to detect both of the final-state particles from the Compton scattering process, a new electromagnetic calorimeter composed of an array of PbWO$_4$ modules was constructed and installed approximately 6~m downstream of the FCAL.
This detector, referred to as the Compton Calorimeter (CCAL), provided angular coverage between 0.2$^{\circ}$ and 0.6$^{\circ}$.
Energy and position of particles detected by the CCAL were reconstructed using a modified version of the \textit{Island} algorithm from GAMS~\cite{Island}, enabling the separation of simply-connected hit patterns into multiple clusters formed by two particles detected in close proximity.
The CCAL was mounted on a movable platform, to be used in special calibration data where the photon beam was directed onto the center of each module.
From these data, the energy resolution for photons detected by the CCAL was determined to be better than 1.5\% for photons with energies of 4~GeV or higher~\cite{Asaturyan2021}.

The main trigger for this experiment was formed by requiring that the energy depositions in the FCAL and CCAL add to at least 3 GeV. 
The offline event-selection criteria consisted of:
\begin{enumerate}
    \item At least one particle detected by the CCAL, at least one particle detected by the FCAL, and at least one beam photon tagged, all with measured times which, after accounting for time of flight, are within $\pm$2~ns of the electron bunch which formed the trigger.
    \item A minimum energy of 3~GeV was required for the CCAL shower, and a minimum energy of 0.5~GeV for the FCAL shower.
    \item Fiducial cuts were applied to remove the regions around the calorimeters that surround the beam hole, where the resolution is poor.
    \item A coplanarity cut was applied on the difference of the azimuthal angles of the two detected particles: $|\phi_\text{CCAL} - \phi_\text{FCAL}| < 5\sigma_{\phi}$, where $\sigma_{\phi}$ is the width of this distribution and is about 6$^{\circ}$.
    \item Energy conservation between the tagged beam photon and the measured energies of the FCAL and CCAL showers was required: $\left|E_{\gamma} - (E_\text{CCAL} + E_\text{FCAL})\right| < 5\sigma_{E}(E_{\gamma})$, where $\sigma_{E}(E_{\gamma})$ ranged from 1.6\% to 2.1\% between $E_{\gamma}$=6~GeV and 11~GeV.
\end{enumerate}
After all cuts were applied, the elasticity distribution:
\begin{equation}
	\Delta K = E_\mathrm{Compton}(\theta_\text{FCAL},\theta_\text{CCAL}) - E_{\gamma}
	\label{eq:deltaK}
\end{equation}
was calculated, where $E_\mathrm{Compton}$ is a function of the measured scattering angles of the particles and represents the beam-photon energy calculated from the kinematics of the two-body Compton scattering process. 
Because the electron mass is small compared to the initial photon energy, the relationship between the scattering angle and energy of the outgoing electron is approximately the same as the photon, meaning the lack of particle identification is unimportant in the calculation of $E_\mathrm{Compton}$.
The data were binned according to the tagged energy $E_{\gamma}$, with separate $\Delta K$ distributions for each bin.

Because of the high beam intensity, often multiple bremsstrahlung photons were tagged within the same 4~ns beam bunch, although only one led to an interaction in the target.
Tagged photons with time correlations corresponding to five beam bunches preceding the prompt photon that coincides with the trigger time, and five bunches following it, were used to estimate the level of the accidental tagger hits within the $\pm$2~ns time window applied to select the beam photon.
We estimated that approximately 8.5\% of tagged beam photons within this timing cut arise from accidental coincidences.

Data collected with the $^{9}$Be target removed were used to study the background originating from interactions of the photon beam outside of the target. 
These data were scaled by the ratio of the photon fluxes collected with and without the $^9$Be target, and were used to subtract this beamline background bin-by-bin from the $\Delta K$ histograms.
The background was estimated to contibute approximately 6\% of all events that survived the event-selection criteria above.

\begin{figure}[ht]
    \centering
    \includegraphics[width=\linewidth]{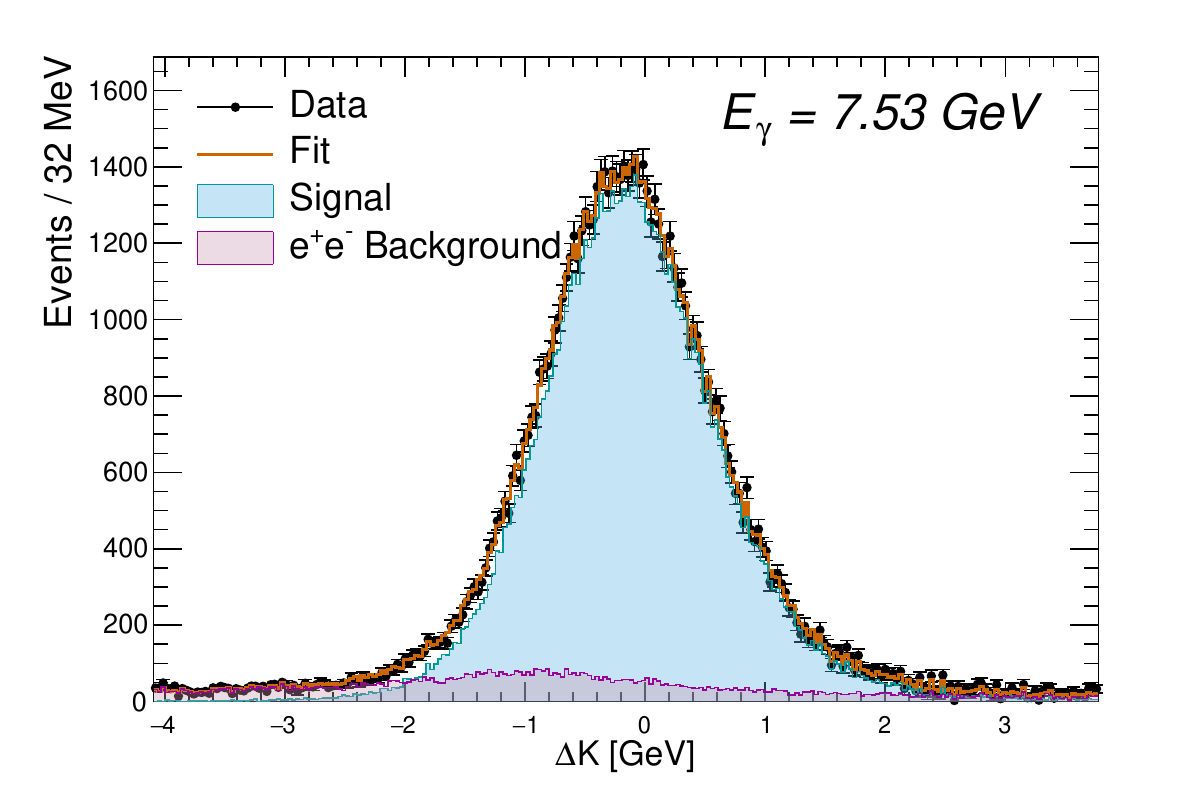}
    \caption{Example of a fit to the $\Delta K$ distribution to extract the Compton yield in one bin of the energy $E_{\gamma}$ of the (tagged) beam photon (here: 7.53~GeV). See Eq.~\ref{eq:deltaK} for the definition of $\Delta K$. Backgrounds from beam-line components and accidentals in the photon tagger have already been subtracted. The remaining background (shown in magenta) is estimated from simulation of $e^{+}e^{-}$ pair production in the target, and the signal line shape (blue) is estimated from the simulation of the Compton scattering signal.}
    \label{fig:yield_fit}
\end{figure}

The black points in Fig.~\ref{fig:yield_fit} represent the $\Delta K$ distribution for one $E_\gamma$ bin after subtraction of the accidental and beamline backgrounds.
The majority of the remaining background originated from $e^{+}e^{-}$ pairs produced from conversions of the beam photons in the $^{9}$Be target.
This process was simulated using a program that generated samples of $e^{+}e^{-}$ pairs weighted according to the differential cross section for Bethe-Heitler production on a $^{9}$Be nucleus, with corrections from incoherent production on quasi-free nucleons, and more importantly from the triplet production process, in which $e^{+}e^{-}$ pairs are produced through interactions of beam photons with the atomic electrons.
The generated $e^{+}e^{-}$ pairs and associated recoil particles ($^{9}$Be for coherent nuclear production, proton or neutron for incoherent production, or $e^{-}$ for triplet production) were propagated through the target and the full experimental setup and the detector response was simulated using a GEANT4-based program~\citep{Allison2016}.
The simulated data were subjected to the same event-selection criteria described above, and used to produce template histograms that were fit to the $\Delta K$ distributions from the experimental data.
An example of the $\Delta K$ distribution for the simulated $e^+e^-$ background is shown for one $E_\gamma$ bin by the magenta histogram in Fig.~\ref{fig:yield_fit}.

In order to estimate the $\Delta K$ line shape of the signal and to calculate the acceptance for the detection of Compton scattering events, the BASES/SPRING-based event generator described in Sec.~\ref{sec:theory} was used to produce a sample of approximately $10^6$ Compton scattering events in each $E_{\gamma}$ bin.
The propagation of the generated particles through the experimental setup and the detector response were simulated using the same GEANT4-based program as for the $e^{+}e^{-}$ background, and the simulated events were analyzed in the same way as the data.
The blue histogram in Fig.~\ref{fig:yield_fit} shows the reconstructed $\Delta K$ distribution for simulated Compton scattering events in one $E_\gamma$ bin.
The peak position of the $\Delta K$ distributions for both experimental data and simulation are offset from 0~GeV due to angular reconstruction biases induced by the clustering algorithms and event selection.
However, the relative offset between the two distributions was consistent, indicating good agreement between data and simulation.
The acceptance was estimated by the fraction of the generated events that remained after applying the selection criteria, and ranged from 11\% at $E_{\gamma}=6.5$~GeV to 4\% at $E_{\gamma}=11$~GeV.

The Compton scattering yield was extracted for each $E_{\gamma}$ bin by fitting the $\Delta K$ distributions with the sum of the signal distribution obtained from the Compton-scattering simulation and the background distribuion obtained from the $e^+e^-$ simulation.
A binned maximum likelihood fit based on Ref.~\cite{Barlow1993} was used to determine the fractional contributions of the two components present in the data, and the Compton yield was taken as the total number of events integrated around a $\pm$5$\sigma$ window around the mean of the $\Delta K$ distribution, multiplied by the fraction found by the fit.
An example of such a fit for one $E_{\gamma}$ bin is shown by the orange histogram in Fig.~\ref{fig:yield_fit}.

\section{Cross-Section Results}

\begin{figure*}[t]
    \centering
    \includegraphics[width=0.85\linewidth]{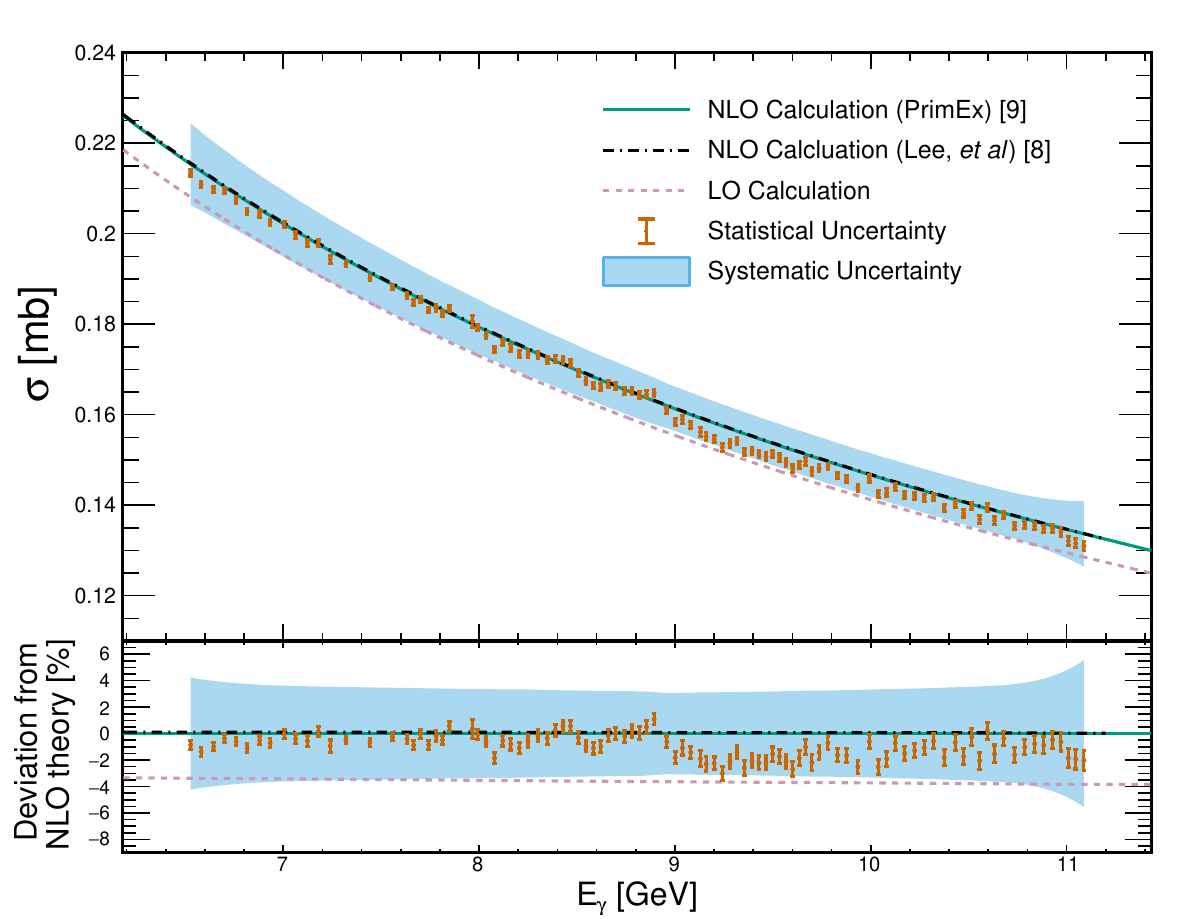}
    \caption{Total Compton scattering cross section measured on atomic electrons of the $^{9}$Be target. The green solid curve corresponds to the NLO calculation performed as part of this work, and the black dash-dotted curve is taken from Ref.~\citep{Lee2021:PRL}. The pink short-dashed line represents the total cross section at LO, obtained by integrating the Klein-Nishina formula in Eq.~\ref{eq:klein-nishina}. Error bars on each point represent statistical uncertainties, whereas the light-blue band, centered at the NLO calculation, represents the total systematic uncertainty. The bottom plot shows the relative difference (in percentage) between the measured values and the NLO calculation.}
    \label{fig:cross_section}
\end{figure*}

The total Compton scattering cross section in each $E_{\gamma}$ bin was obtained by normalizing the measured Compton yields by the luminosity and the acceptance.
The results are shown in the top panel of Fig.~\ref{fig:cross_section}, where each point is plotted with error bars representing the statistical uncertainty. 
The results of the numerical integration of the total cross section at NLO (see Sec.~\ref{sec:theory}) is represented by the solid green curve, and the calculation using a recently-obtained analytical expression for the NLO cross section~\cite{Lee2021:PRL} is represented by the dash-dotted black curve. 
As stated in Sec.~\ref{sec:theory}, the two calculations agree to better than 0.1\%.
The LO cross section obtained by integrating the Klein-Nishina formula in Eq.~\ref{eq:klein-nishina} over all photon angles (pink short-dashed curve in Fig.~\ref{fig:cross_section}) is 3.5\% to 4\% lower than the NLO value.
Finally, the total systematic uncertainty on the experimental results is indicated by the light-blue band, and is centered at the NLO calculation.

The bottom part of Fig.~\ref{fig:cross_section} shows the deviation (in percentage) of the measured cross-section values from the NLO calculation in Ref.~\cite{PrimExCompton}.
In all $E_{\gamma}$ bins, the measured cross section is consistent with the NLO calculation within the estimated uncertainty. 
The statistical uncertainty ranges from 0.3\% in the lowest $E_{\gamma}$ bins to 1.0\% in the highest $E_{\gamma}$ bins, which reflects the decrease of the total cross section and the acceptance as a function of $E_{\gamma}$.
While the agreement between the data and the NLO calculation is good in general considering the overall experimental uncertainties, we note a slightly different trend between the two above 9~GeV.
This trend may indicate a systematic effect, but its possible origin is not addressed within this article.

\begin{table}[ht]
    \centering
    \caption{Estimated systematic uncertainties in percentage of the measured total Compton scattering cross section for three representative energy bins. Rows in bold represent broad categories of systematic uncertainty, and (where applicable) represent the quadrature sum of the plain-text items listed beneath them.}
    \begin{adjustbox}{max width=\linewidth}
    \begin{tabular}{|l c c c|}
    \hline
       Source of Uncertainty  & $E_{\gamma}=7.1$~GeV & $E_{\gamma}=8.4$~GeV & $E_{\gamma}=10.0$~GeV \\
       \hline
       \hline
        \multicolumn{1}{|c}{\textit{Energy-Dependent}} & & & \\
        \textbf{Photon Flux} & \textbf{2.18} & \textbf{1.65} & \textbf{1.68} \BSTR \\ 
        \textbf{Event Selection} & \textbf{1.50} & \textbf{1.38} & \textbf{1.42} \\
        \hspace{3mm} Elasticity Cut & 0.59 & 0.68 & 0.79 \\
        \hspace{3mm} Detector Fiducialization & 0.42 & 0.54 & 0.74 \\
        \hspace{3mm} Minimum Energy Cuts & 0.28 & 0.17 & 0.17 \BSTR \\
        \hspace{3mm} $\Delta K$ Cut & 1.28 & 1.05 & 0.89 \\
        \textbf{Geometric Acceptance} & \textbf{0.45} & \textbf{0.53} & \textbf{0.73} \BSTR \\
        \textbf{Background Subtraction} & \textbf{0.64} & \textbf{0.82} & \textbf{0.73} \\ 
        \hspace{3mm} Accidental Background & 0.25 & 0.22 & 0.20 \\
        \hspace{3mm} Beam-line Background & 0.34 & 0.30 & 0.25 \\
        \hspace{3mm} $e^+e^-$ Pair Background & 0.48 & 0.73 & 0.66 \\ \hline \hline
        \multicolumn{1}{|c}{\textit{Normalization}} & & & \\
        \textbf{Target} & \textbf{0.25} & \textbf{0.25} & \textbf{0.25} \\
        \textbf{Detector Efficiency} & \textbf{2.24} & \textbf{2.24} & \textbf{2.24} \\ \hline
       \hline
       \textbf{Total} & \textbf{3.56} & \textbf{3.27} & \textbf{3.31} \\
       \hline
    \end{tabular}
    \end{adjustbox}
    \label{tab:sys_uncertainties}
\end{table}

A breakdown of the relative systematic uncertainties for 3 representative $E_{\gamma}$ bins is given in Table~\ref{tab:sys_uncertainties}.
The top part of Table~\ref{tab:sys_uncertainties} lists the sources of systematic uncertainty that vary with the photon beam energy. 
The largest individual contribution to the energy-dependent systematic uncertainty comes from the photon flux measurement, and is primarily due to uncertainties in the shape of the PS acceptance.
Due to the finite statistics available in the PS calibration data, the acceptance was parameterized using analytical functions, with the shapes determined by fits to the calibration data. 
Energy-dependent systematic uncertainties on the acceptance shape were estimated by varying the fit parameters by $\pm 1\sigma$ and evaluating the resulting variations in the acceptance curves. 
Further uncertainty was assigned to account for differences arising from alternative parameterizations of the acceptance function. 
The combined relative uncertainty from these sources ranged between 1.3\% and 1.5\% in the photon energy range of 7.1~GeV to 10~GeV. 
Additionally, a conservative systematic uncertainty was assigned to account for potential long-term variations, based on comparisons of different PS calibration runs and evaluations of the stability of the detector over time.
Other sources of systematic uncertainty were investigated, including beamline-related backgrounds and contributions associated with the subtraction of accidental hits in the PS detectors. 
These effects were found to be negligible ($\ll$1\%). 
All sources of systematic uncertainty were combined in quadrature to determine the total uncertainty on the photon flux measurement.
Moreover, because the PS acceptance gradually drops to nearly zero below 6.5~GeV and above 11.1~GeV, the uncertainty on the photon flux in those regions is significantly larger.
This causes the widening of the systematic uncertainty band in Fig.~\ref{fig:cross_section} near the edges of the measured energy range.

Systematic uncertainties related to the accuracy of the detector simulation were estimated by varying all of the event-selection criteria independently and recording the change in the total cross-section measurement.
The largest effect was observed to come from the $\Delta K$ fit range used to extract the signal yield.
We observed a change of 1\% in the cross-section result when varying the $\Delta K$ fit range from the nominal value of $\pm5\sigma$ around the mean, to either $\pm3\sigma$ or $\pm10\sigma$ around the mean.
When combined in quadrature, the systematic uncertainties estimated from the cut variations were approximately 1.5\%.

The uncertainty due to the geometric acceptance was studied by artificially altering the positions of the FCAL and CCAL with respect to the nominal photon beam axis in the simulation.
While inaccuracies of the experimental geometry in the simulation would lead to relatively large systematic effects in the differential cross section as a function of the photon scattering angle, their effect on the total cross-section measurement is less significant.

Systematic uncertainties on the subtraction of the accidental background as well as the non-target background were estimated by the change in the experimental Compton yield when their relative contributions were artificially increased or decreased within their known accuracies. 
The systematic uncertainty on the $e^{+}e^{-}$ background subtraction was studied by altering the shape of the template histogram used for the yield-extraction fit.

The bottom part of Table~\ref{tab:sys_uncertainties} lists sources of systematic uncertainty that contribute equally across all beam energies (normalization uncertainties).
The uncertainties on the target thickness and purity contribute only 0.25\%.
The systematic uncertainties related to the accuracy of the FCAL and CCAL detection efficiencies are larger, however.
For the FCAL, we estimated a systematic uncertainty of $\pm$2\% based on comparisons of detection efficiencies of photons from $\omega\rightarrow\pi^+\pi^-\pi^0(\gamma\gamma)$ decays between data and MC.
Finally, special calibration data were collected by centering each module of the CCAL directly in the path of a low-intensity photon beam.
These data were used to compare the measured and simulated CCAL energy response functions, from which we estimated an additional 1\% systematic uncertainty.

\section{Conclusions}

We measured the total cross section for Compton scattering of photons on atomic electrons in $^{9}$Be atoms, using the GlueX spectrometer in Hall D at Jefferson Lab achieving a precision of 3.4\% averaged over all $E_{\gamma}$ bins when adding the statistical and systematic uncertainties in quadrature.
This measurement provides the first experimental result of the total cross section for photon energies above 6~GeV, and shows closer agreement with QED calculations that include NLO corrections.
Additionally, this is the first high-precision absolute cross-section measurement conducted using the GlueX spectrometer and photon tagging system, serving as a crucial validation of the experimental setup for future cross-section measurements at forward angles.

\section*{Acknowledgements}

This work was supported in part by the U.S. Department of Energy, Office of Science, Office of Nuclear Physics under Contract No. DE-FG02-03ER41231, as well as NSF grants PHY-1812396, PHY-2111181, and PHY-2412800.

We would like to acknowledge the outstanding efforts of the staff of the Accelerator and the Physics Divisions at Jefferson Lab that made the experiment possible. This work was supported in part by the U.S. Department of Energy, the U.S. National Science Foundation, the German Research Foundation, GSI Helmholtzzentrum f\"ur Schwerionenforschung GmbH, the Natural Sciences and Engineering Research Council of Canada, the Russian Foundation for Basic Research, the UK Science and Technology Facilities Council, the National Natural Science Foundation of China and the China Scholarship Council. This material is based upon work supported by the U.S. Department of Energy, Office of Science, Office of Nuclear Physics under contract DE-AC05-06OR23177.

\bibliographystyle{elsarticle-num-names}
\raggedright
\bibliography{main}

\end{document}